\documentclass{aa}
\usepackage{graphicx}
\usepackage{txfonts}
\usepackage{natbib}
\bibpunct{(}{)}{;}{a}{}{,}

\def\la{\mathrel{\hbox{\rlap{\hbox{\lower4pt\hbox{$\sim$}}}\hbox{$<$}}}}
\def\ga{\mathrel{\hbox{\rlap{\hbox{\lower4pt\hbox{$\sim$}}}\hbox{$>$}}}}

\newcommand{\msol}{\ensuremath{{\mbox{M}_\odot}}}
\newcommand{\kmps}{\ensuremath{\mbox{km}~\mbox{s}^{-1}}}

\newcommand{\alambda}{ \left[ \frac{\beta (1 -\mu'^2)}{r} +
\gamma^2 \mu' (\mu' + \beta ) \frac{\partial \beta }{\partial r}      \right]}

\begin{document}
\title{
Analyzing \object{SN 2003Z}\thanks{Based on observations collected at the Centro Astronómico Hispano Alemán (CAHA) at Calar Alto, operated jointly by the
Max-Planck Institut für Astronomie and the Instituto de Astrofísica de Andalucía (CSIC)} 
with {\tt PHOENIX} 
 }
\author{Sebastian Knop \inst{1} \and Peter H. Hauschildt \inst{1} \and
  E. Baron \inst{1,2,3} \and Stefan Dreizler \inst{4} }
\institute{
Hamburger Sternwarte, Gojenbergsweg 112, 21029 Hamburg, Germany \\ email:~\texttt{sknop@hs.uni-hamburg.de;yeti@hs.uni-hamburg.de} \and
University of Oklahoma, 440 West Brooks St, Rm 100, Norman, OK 73019, USA \\ email:~\texttt{baron@nhn.ou.edu}
\and Computational Research Division, Lawrence Berkeley
  National Laboratory, MS 50F-1650, 1 Cyclotron Rd, Berkeley, CA
  94720-8139 USA
\and 
Institut f\"ur Astrophysik, University of G\"ottingen, Friedrich-Hund-Platz 1, 37077 G\"ottingen, Germany\\ email:~\texttt{dreizler@astro.physik.uni-goettingen.de}}
\date{Received \ Accepted }

\abstract
{}
{We present synthetic spectra around maximum for the type II supernova SN 2003Z,
which was first detected on January 29.7 2003. Comparison with observed spectra 
aim at the determination of physical parameters for SN 2003Z.}
{Synthetic spectra are calculated with our stellar atmosphere 
code \texttt{PHOENIX}.
It solves the special relativistic equation of radiative transfer, 
including large NLTE-calculations and line blanketing by design, 
in 1-dimensional spherical symmetry.
The observed spectra were obtained at the 3.5 meter telescope at
Calar Alto. The TWIN instrument was used so that a spectral range from about
3600 to 7500 {\AA} was covered.  The spectra were taken on Feb.\ 4, 5, 9, and
11, 2003.  }
{The physical parameters of the models give the luminosities, 
a range of possible velocity profiles for the SN, an estimate of the 
colour excess, and the observed metalicity.}
{}


\maketitle

\section{Introduction}

Type II supernovae (SNe II) are thought to originate from the core collapse
of massive stars. By definition SNe II show strong Balmer lines in
their spectra, and thus are thought to be from stars with much of
their hydrogen envelopes intact at the time of the explosion. Largely
through the use of \emph{HST}, progenitors have been identified for
several SNe II, and they seem to come from rather low mass stars $9
\ga M \la 12$~\msol \,\citep{maund05cs05,maund05a,smartt01du03,smartt03gd04,vandyk03a,vandyk01du03,vandyk03gd03}. 

Since the spectra of SNe II form in hydrogen dominated atmospheres
with relatively simple
density structures, they should be among the most accurate spectra to
model with detailed model atmosphere codes. 

Indeed, SNe II can be modeled in detail and reddening, primordial metalicity, and even
distances can be determined
\citep{bsn99em00,sn93w103,bsn99em04}. Thus, through detailed spectral
modeling we can determine important physical parameters that can be
compared with stellar evolution and explosion modeling.

Here, we use the \texttt{PHOENIX} relativistic model atmosphere code package
\citep{jcam,peter1997ApJ,peter1998ApJ,peter2001ApJ} to
simulate the SN II atmosphere for SN 2003Z during its optically thick
(photospheric) phase. 

In the following we first describe the observations of SN 2003Z, and
then describe 
our modeling process including the assumptions made to construct
a model. After that we present our results about luminosities, extinction,
velocity profiles. Finally the results are summarized in the conclusion.

\section{The supernova SN 2003Z}

SN 2003Z was discovered on January 29, 2003 \citep{sn2003za} 
and is a type II supernova \citep{sn2003zb}.

The SN was observed on the 4th, 5th, 9th, and 11th of February 2003 with the 3.5 meter telescope at Calar
Alto. The TWIN spectrograph was used, covering the spectral range from
about 3600 to 7400 {\AA}.  The SN is located near the galaxy \object{NGC 2742}
and is believed to lie in the outer part of a spiral arm of the
galaxy, typical for a Type II supernova.  The redshift of NGC 2742 is
$z=0.0043$ \citep{UZC}.  

From NED\footnote{This research has made use
  of the NASA/IPAC Extragalactic Database (NED) which is operated by
  the Jet Propulsion Laboratory, California Institute of Technology,
  under contract with the National Aeronautics and Space
  Administration.} the foreground colour excess of the galaxy is
$E(B-V)=0.039$ \citep{schlegel}. Information about the
radial velocity of the SN itself or the colour excess of the host galaxy
are unavailable.  

\begin{figure}
\begin{center}
\includegraphics[width=\hsize]{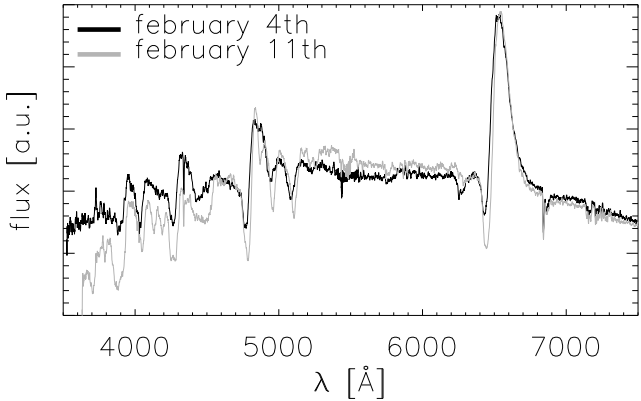}
\caption{Comparison of the spectra of the first and last day of the
observations. A shift in wavelength towards the red and a decrease of flux in
the blue part are visible, indicating the cooling and slowing of the material
producing the spectra.}
\label{fig0}
\end{center}
\end{figure}

The physical structure of the atmosphere changes
over time due to the expansion and the gradual cooling of the ejected
material.  Therefore, the spectrum also changes over time.  From day
to day the changes are small; however, the spectra of the first and
the last day of the observations show large differences as displayed in
Fig.~\ref{fig0}.

The gradual change in the wavelength shift of the spectral
lines originates from the changing optical depth of the cooling and
expanding envelope. As time progresses we see into slower and ``deeper''
layers of the atmosphere so that the lines show smaller redshifts in
the later spectra.

\section{Modeling}

\texttt{PHOENIX} is a general purpose NLTE stellar atmosphere code
package.  For the calculation of the radiative transfer with 
\texttt{PHOENIX} the structure of the atmosphere is approximated by a spherically
symmetric shell, thereby reducing the system to one spatial
dimension. Furthermore, the envelope is assumed to be homologously
expanding, giving a linearly increasing velocity field (Sedov
solution)
\begin{equation}
\label{velocitylaw}
v (r) = v_0 \frac{r}{R_0}
\end{equation}
with $v_0$ being the velocity and $R_0$  the radius at $\tau = 1$ in the radial optical depth grid. 
The run of the density is approximated by a power law \citep{zeldovich67}
\begin{eqnarray}
\rho(r) & = & \left( \frac{r}{R_0} \right)^{-n} \label{rhopowerlaw}
\end{eqnarray}

Due to the fact that the atmosphere of a SN is expanding at
$\simeq 10$ percent of the speed of light, the special relativistic
equation \citep{mihalas84} of radiative transfer in moving media must
be solved.  In co-moving form written in terms of wavelengths we have
\begin{eqnarray}
\eta_\lambda  - \chi_\lambda I_\lambda& = & \frac{\partial I_\lambda}{\partial s}  
+ a_\lambda \frac{\partial \lambda I_\lambda}{\partial \lambda} 
+ 4 a_\lambda I_\lambda
 \label{RTeq} \\
{\rm with:} \quad a_\lambda & = & \gamma \alambda \nonumber
\end{eqnarray}
where $I_\lambda$ is the specific intensity and the differentiation of
the path-length $s$ is the differentiation along the monochromatic path
of a photon in the co-moving frame. Furthermore, $\eta_\lambda$ is the
emissivity and $\chi_\lambda$ the extinction coefficient.  The
emissivity $\eta$ depends via scattering on the mean intensity
$J_\lambda$ itself, therefore, the problem is not a mere differential
equation but an integro-differential equation.  The radiative transfer
problem is solved with an operator splitting technique in the
co-moving frame \citep{petereddie2004,peter1992JQSRT}. For the NLTE
models, the coupled radiation transport and rate equation problem is
solved by a multi-level operator splitting method
\citep{peter1993JQSRT}. 

Details of the code and the numerical methods used in \texttt{PHOENIX} can be found in \citet{jcam}.

All models were calculated on a radial grid with 100 layers, using a
logarithmic optical depth grid ($\tau$-grid) in the continuum at 5000 {\AA}
with a range from $\tau = 10^3-10^{-10}$. 

Other model parameters vary from model to model. For instance
the final model for the first day of the observation had a
pressure at $\tau =1$ of $4.44 \cdot 10^{-8}$ bar and the
corresponding radius was $9.62 \cdot 10^{14}$ cm.

Unstable nuclei -- most importantly \element[][56]{Ni} -- are created in SN outburst and, therefore, the input
of radioactive energy plays an important role in the formation of the
spectrum.  The treatment of the influence of radioactivity is
simplified by the assumption that $\gamma$-ray deposition follows the
density profile and the total mass of radiative nuclei is a parameter
in the modeling.
In all models we iterated the temperature structure to fulfill the
condition of energy conservation in the co-moving frame. Furthermore,
the most important species were treated in complete NLTE. 
(see Table ~\ref{table1}). The decision which ionization stages were included in
NLTE was made by taking into account the partial pressures of those
stages. An ion was ignored for the NLTE calculation only if the
pressure of an ion was for all layers was at least 15 dex 
smaller than the pressure of the dominant ion of the species.

To model SN atmospheres, a few physical parameters must be
specified. For example, the luminosity, the
exponent $n$ of the density power law (\ref{rhopowerlaw}), and the
velocity field of the atmosphere must be specified. These parameters
are not known in advance. 
Therefore, several models with different physical parameters had to be calculated and the
best set of parameters was determined by comparison of the observed
and synthetic spectra.  The first observation was modeled to determine
the basic parameters of the SN and the following observations were
reproduced by changing the physical parameters appropriately.

\section{Results}

To compare the model spectra with the observation a couple of effects
have to be taken into account, such as the redshift of the SN, 
and extinction.

Nothing is known about the intra-galactic extinction in NGC 2742, hence we
only correct for the known foreground extinction of NGC 2742 $E(B-V)=0.039$.

We correct the model spectra for the redshift $z=0.0043$ of the host galaxy and
for the wavelength shift due to the the transition from vacuum to
air\footnote{This 
correction is very small and we apply it only for completeness.}.

\begin{figure}
\begin{center}
\includegraphics[width=\hsize]{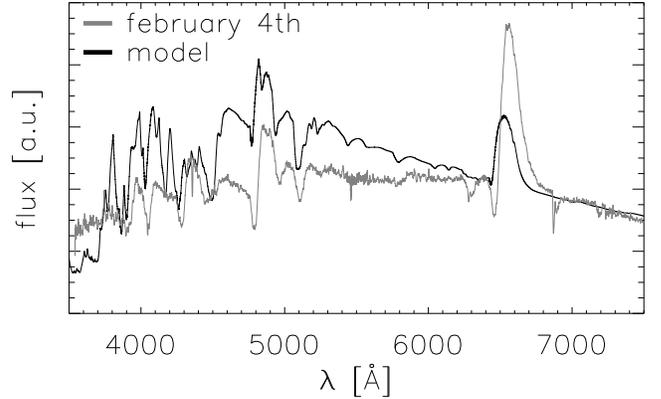}
\caption{The plot shows the first model that reproduced some of the
  observed spectral features. No model was found that matched the
  observed rise of the continuum, hence the extinction inside the
  parent galaxy appears to be important. 
  The spectra were normalized in such a way that they
  match the continuum above 6900 {\AA}.}
\label{fig1}
\end{center}
\end{figure}

First we ran a couple of models varying the luminosity (model temperature)
and the velocity field to get a first guess for those basic parameters.
After including the first four
ionization stages of iron as well as hydrogen and helium in complete
NLTE, a spectrum was obtained that fits some of the key features of
the observed spectrum, see Fig.~\ref{fig1}. The displayed model spectrum
has $T_\mathrm{model} = 5600 \mathrm{K}$, a density exponent $n=11$ 
and $v_0=6300$ \kmps (see Eq. \ref{velocitylaw}). The model spectrum is 
blue shifted with respect to the observation. This is due to difficulties
with the convergence of the models that had lower photospheric 
velocities. 
This initial result is clearly due to a velocity that was too large in the
models. This simple preliminary model had few species in NLTE and it was
difficult to converge models with lower velocities. As we improved the input
physics in our calculations, this problem did not persist.

The most obvious flaw besides the wrong velocity, is the steepness of the
continuum in the wavelength range from 5200 to 6200 {\AA}. 
After calculating a grid of models it became clear that there is no
combination of the model parameters that reproduces this slope of
the continuum.

\begin{figure}
\begin{center}
\includegraphics[width=\hsize]{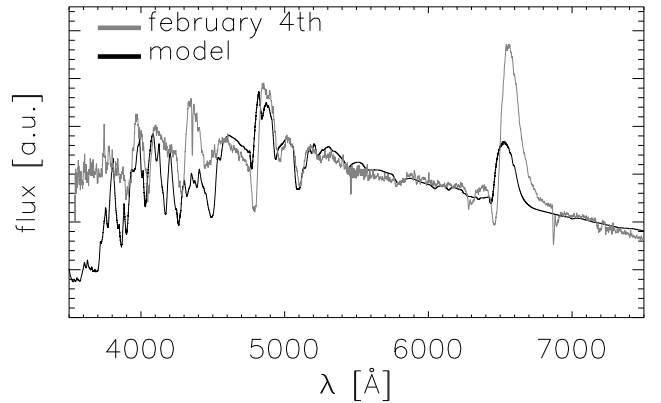}
\caption{The same model as in Fig.~\ref{fig1} is shown. The 
  observation was dereddened using a higher extinction of
  $E(B-V)=0.3$.  The spectra were normalized to match
  the continuum above 6900 {\AA}.}
\label{fig2}
\end{center}
\end{figure}

However, it is plausible and to be expected that there is additional 
extinction present in the host galaxy. Therefore, we correct the observation
for additional reddening. With an assumed extinction of $E(B-V)=0.3$,
the continuum of the model from 
Fig.~\ref{fig1} was matched very well (See Fig.~\ref{fig2}).
Since the assumed value of extinction is an arbitrary choice and is only
a lower boundary for possible values of the extinction we 
ran a number of tests to place limits on the extinction coefficient.  Large
values of $E(B-V)$ demand high model temperatures, but higher
temperatures yield synthetic spectra that are dominated by the Balmer
series of hydrogen and other spectral features were diminished.

Hence we decided to choose the extinction as small as needed to match
the model. During the progress of the analysis,
due to extensive NLTE calculations, non-solar abundances and
taking the other observed epochs into account. In the final
model the extinction was chosen to $E(B-V)=0.4$.
Since the continuum wasn't a reliable temperature indicator
anymore we had to rely on reproduced spectral details to determine
the model temperatures.

Because the model spectrum in Fig.~\ref{fig1} was very promising, we
have based our further modeling on the parameters of this
model. It already includes hydrogen, helium and the first four ionization stages of iron
in NLTE. During the course of the modeling we included all the species
in Table~\ref{table1} in NLTE. 
Further extensive model grids were calculated, varying the model
temperature, the density exponent, time since explosion (important
for the radioactive decays), and the velocity field.

\begin{table}[h!]
\caption{This table lists all ionization stages that were treated in NLTE.}
\label{table1}
\centering
\begin{tabular}{ r  l  r  l  }
\hline
\hline
Z & Ions &  Z  & Ions  \\ \hline 
1 & \ion{H}{i} &  14 & \ion{Si}{i-iii} \\
2 & \ion{He}{i}  & 15  & \ion{P}{i-ii} \\
6 & \ion{C}{i-ii} &  16  & \ion{S}{i-ii} \\
7 & \ion{N}{i-ii} & 19  & \ion{K}{i-ii} \\
8 & \ion{O}{i-ii} & 20  & \ion{Ca}{i-iii} \\
10 & \ion{Ne}{i} & 25  & \ion{Mn}{i-ii} \\
11 & \ion{Na}{i-ii} & 26  & \ion{Fe}{i-iv} \\
12 & \ion{Mg}{i-iii} & 27  & \ion{Co}{i-iii} \\
13 & \ion{Al}{i-ii} & 28  & \ion{Ni}{i-iv}  \\ \hline
\end{tabular}\\
\end{table}

We found that the spectrum of Feb.~4th was best fitted with
an effective temperature of 5800\,K and a density exponent $n=9$.  The
exact age of the SN proved to be not particularly important.  However,
it is clear that the radioactivity is necessary to reproduce the
features in the observed spectra as it influences the temperature structure.  

To further improve the model quality we dropped the assumption of 
solar abundances and did an abundances analysis.
Assumed under abundances of specific species such as iron, 
carbon, or nitrogen of up to [Z/H]$=-6$ resulted in little or no effects on the
spectra. The most noticeable effect resulted from an overall
under abundance of all the metals. The spectra fit significantly better if 
the metalicity was [M/H]$=\log{\frac{3}{10}}\approx-0.5$. However, since 
the fit is not good enough to use quantitative methods this is only determined by eye. 
 Lower metalicities seem to fit the observation for a broader range of values.
Hence the progenitor seems to have been  a metal-poor star.


\begin{figure}
\begin{center}
\includegraphics[width=\hsize]{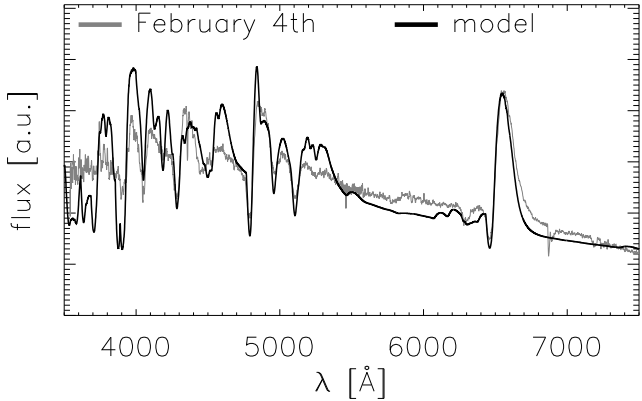}
\caption{A model is compared to the observation on Feb.~4.  Solar
  abundances are assumed.  The other key-parameters are
  $T_{\rm{model}}=5800$ K and $v_0=4900$ \kmps.  The observation was
  corrected for $E(B-V)=0.4$. The spectra were normalized by
  area.}
\label{fig4}
\end{center}
\end{figure}
\begin{figure}
\begin{center}
\includegraphics[width=\hsize]{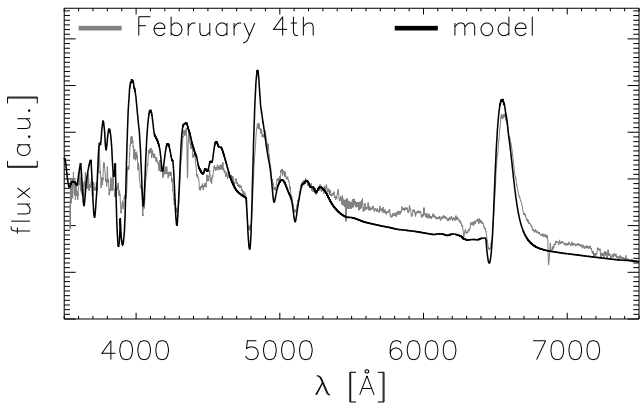}
\caption{The best fitting synthetic spectrum
  is compared to the observation on Feb.~4.  All model
  parameters are the same as in Fig.~\ref{fig4} with the exception of
  the metalicity which is [M/H]$\approx -0.5$.  The observation was 
  corrected for $E(B-V)=0.4$. In comparison with the model
  from Fig.~\ref{fig4} the fit is now much better. The spectra were
  normalized by area.}
\label{fig5}
\end{center}
\end{figure}

The variation of the velocity field was required in order for the
models to match the observations.  
Gradually changing the velocity field
resulted in a different wavelength shift and only slightly changing spectral
features. No particular velocity could be favoured other another in a range of a few 
hundred \kmps. 

The variation of the
velocity was problematic in the first models due to the fact that slower models
didn't converge properly or didn't reproduce the spectra as well as before. However, 
due to the refining of the atmosphere structure via inclusion of more NLTE species
the models became more robust to the change of the velocity parameter.

\begin{figure}
\begin{center}
\includegraphics[width=\hsize]{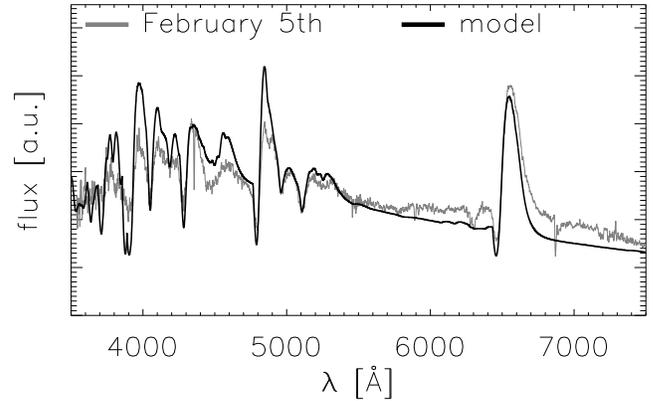}
\caption{The best fitting synthetic spectrum for Feb.~5 is compared to
  the observation. The model has $T_{\rm{model}}=5700$ and $v_0 = 4850
  \kmps$. The spectra were normalized by area.}
\label{fig8}
\end{center}
\end{figure}

\begin{figure}
\begin{center}
\includegraphics[width=\hsize]{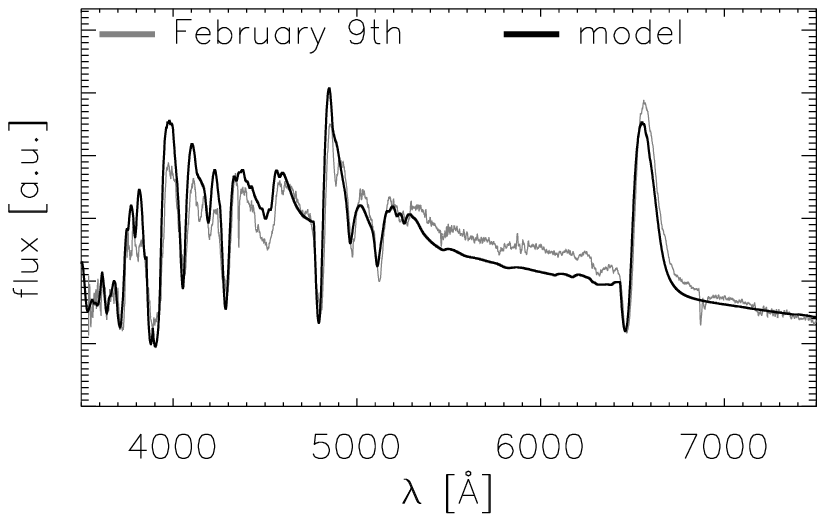}
\caption{The best fitting synthetic spectrum for Feb.~9 is compared to
  the observation. The model has $T_{\rm{model}}=5600$ and $v_0 = 4500
  \kmps$. The spectra were normalized by area.}
\label{fig9}
\end{center}
\end{figure}

\begin{figure}
\begin{center}
\includegraphics[width=\hsize]{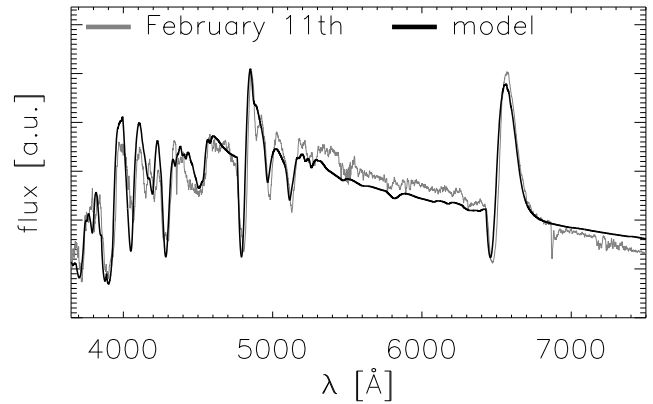}
\caption{The best fitting synthetic spectrum for Feb.~11 is compared to
  the observation. The model has $T_{\rm{model}}=5400$ and $v_0 = 4300
  \kmps$. The spectra were normalized by area.}
\label{fig10}
\end{center}
\end{figure}

In Figs. ~\ref{fig5}--\ref{fig10} we show the final model spectra
for all four observed epochs. In all Figs. the observations were
corrected for an extinction of $E(B-V)=0.4$. This change of the extinction
from the initial guess is mainly due to the use of non-solar abundances. 
The under-abundance of the
metals increased the flux in the blue part of the spectrum and therefore the
observations demanded more de-reddening. In order to support
  the large reddening that we find, we have searched for Na~ID
  interstellar absorption lines
  in the host galaxy. Since there is a strong broad SN feature just as
  the wavelength of the Na~ID lines they don't clearly stand out, but
  there is a hint of an extra narrow absorption line at the right
  wavelength. Thus, our somewhat high extinction value is reasonable.

The important model parameters of the different observations are
summarized in Table \ref{table2}.


\section{Conclusion}

We have modeled the early spectra of the Type II supernova SN 2003Z.
The results show that there is substantial extragalactic reddening
towards the SN, probably due to dust in the plane of its parent
galaxy. For each observed epoch we have determined the best-fit
parameters of the NLTE models as summarized in Table~\ref{table2}. 

The best fits
require sub-solar metalicities of about [M/H]$\approx -0.5$, indicating
that the progenitor of SN 2003Z was a quite metal-poor star (see also
\cite{sn93w103}). 
Unfortunately there are only very few observations available for this
object and in particular, no accurate light curve is known,
therefore we have assumed that the relative fluxes are
  accurate. Without photometry we can not calibrate the relative
  fluxes and thus our results are sensitive to the accuracy of the
  relative flux calibration.

\begin{table}
\caption{Summary of the basic parameters $T_{\rm{model}}$ and velocity for the
  different days. There is no error estimation for the velocity as a range of
  velocities fit the spectra 
  and the given value was just picked to match the
  observed redshift. The models fit best with metalicities [M/H]$\approx -0.5$, however,
  there is quite a range of possible values that cannot be ruled completely out, so
  [M/H]$= -0.5 +/- 0.3/0.5$. The
  luminosities are relatively well determined -- $\Delta T_{\rm{model}}=\pm 100 \mathrm{K} $ -- for the assumed extinction, but depend
  strongly on this parameter. The pressure $p_0$ is determined at $\tau =1$ in the radial $\tau$-grid.}
  \label{table2}
  \centering
\begin{tabular}{l l l l l l}
\hline
\hline
February & $T_{\mathrm{model}}$ & [M/H] & $v_0 $ & n & $p_0 $ \\
         &    [K]         &       & [\kmps]  &   & $[10^{-8} \,\mathrm{bar}]$\\                     
\hline
4th  & 5800   & -0.5  & 4900  & 9 & 4.43 \\
5th  & 5700   & -0.5  & 4850  & 9 & 5.41 \\
9th  & 5600   & -0.5  & 4500  & 9 & 6.83 \\
11th & 5400   & -0.5  & 4300  & 9 & 9.76 \\
\hline
\end{tabular}
\end{table}
\begin{acknowledgements}
This work was supported in part  by NASA grants
NAG5-3505 and NAG5-12127, and NSF grants AST-0204771 and AST-0307323, 
PHH was supported in part by the P\^ole Scientifique de Mod\'elisation
Num\'erique at ENS-Lyon. Some of the calculations presented here were
performed at the H\"ochstleistungs Rechenzentrum Nord (HLRN), and at
the National Energy Research Supercomputer Center (NERSC)
which is supported by the Office of Science of the U.S.  Department of
Energy under contract DE-AC03-76SF00098.
We thank all these institutions for a generous
allocation of computer time.
\end{acknowledgements}

\bibliographystyle{aa}

\end{document}